\documentclass{article}

\usepackage[a4paper, margin=25mm]{geometry}
\usepackage[colorlinks, allcolors=blue]{hyperref}
\usepackage[backend=biber, style=ieee]{biblatex}
\usepackage{url}

\usepackage{graphicx}
\usepackage{booktabs}
\usepackage{multicol}

\title{\huge\bf Cyber~Threat~Intelligence for Artificial~Intelligence~Systems}
\author{
    Natalia Krawczyk, Mateusz Szczepkowski, Adrian Brodzik, and Krzysztof Bocianiak \\\\
    \it Orange Innovation Poland
}
\date{}

\begin{document}

\maketitle

\begin{abstract}
As artificial intelligence (AI) becomes deeply embedded in critical services and everyday products, it is increasingly exposed to security threats which traditional cyber defenses were not designed to handle. In this paper, we investigate how cyber threat intelligence (CTI) may evolve to address attacks that target AI systems. We first analyze the assumptions and workflows of conventional threat intelligence with the needs of AI-focused defense, highlighting AI-specific assets and vulnerabilities. We then review and organize the current landscape of AI security knowledge. Based on this, we outline what an AI-oriented threat intelligence knowledge base should contain, describing concrete indicators of compromise (IoC) for different AI supply-chain phases and artifacts, and showing how such a knowledge base could support security tools. Finally, we discuss techniques for measuring similarity between collected indicators and newly observed AI artifacts. The review reveals gaps and quality issues in existing resources and identifies potential future research directions toward a practical threat intelligence framework tailored to AI.
\end{abstract}

\begin{multicols}{2}

\section{Introduction}

Cyber attacks such as ransomware, phishing, and social engineering have increased in recent years, forcing both companies and public institutions to pay more attention to security~\cite{aldauiji2022utilizing}. Getting exact numbers on global cyber attacks is difficult, as many go unreported and attack definitions may vary; but recent research gives us good estimates and shows clear trends. For instance, one study found 11,497 cybersecurity incidents worldwide from October 2023 to March 2025, covering 106 attack types across 257 countries~\cite{sufi2025quantifying}. Additionally, forecasts predict 1,782-2,080 incidents per quarter through 2026, with nearly half affecting multiple countries.

Another dataset~\cite{sufi2024new} collected 77,623 cyber attack records from 225 countries over 14 months (Oct 2022-Dec 2023) using AI data collection methods. It includes various attack types like spam, ransomware, exploits, and web threats, and shows how different threats appear around the world.

Beyond the growing number of attacks, their impact is getting worse. Estimates suggest cyber crime could cost around \$10.5 trillion per year by 2025, up from roughly \$3 trillion in 2015~\cite{kazim2024impact}. The most affected sectors include finance, energy, and healthcare, where attacks can interrupt essential services, expose sensitive data, and generate both financial losses and reputational damage. These trends push organisations to strengthen security and invest in better protection measures.

As cyber threats grow more complex, organizations increasingly rely on cyber threat intelligence (CTI) to better understand, detect, and prepare for attacks. CTI refers to proactive identification and analysis of cyber threats. It is the process of collecting, analyzing, and sharing information about cyber threats to help organizations anticipate, detect, and respond to cyber attacks~\cite{alaeifar2024current}. CTI includes information about threat actors, their tactics, techniques, and procedures (TTPs), vulnerabilities, and indicators of compromise (IOCs)~\cite{alaeifar2024current}.

As artificial intelligence (AI) technology evolves quickly, the threat landscape is evolving as well. AI systems create new attack surfaces, from traditional machine learning (ML) models to large generative models. This gives attackers new ways to develop attack techniques that do not fit existing CTI categories. Recent empirical studies and security databases highlight the significant scale of attacks on AI and ML systems. For example, an analysis using the MITRE ATLAS AI Model Vulnerabilities Dataset recorded 500 adversarial attacks, with 53.6\% targeting model-level vulnerabilities, 35.2\% data-level, and 11.2\% deployment-level~\cite{olutimehin2025adversarial}. Another study documented 89 real-world ML attack scenarios from the MITRE ATLAS and AI Incident Database, showing that convolutional neural networks (CNNs) are among the most attacked models~\cite{tidjon2022threat}. These numbers represent only a subset of reported incidents, as many attacks go unreported or are discovered post-deployment.

This creates a need for CTI frameworks designed for AI-related threats. Such frameworks would require defining new types of IoCs, and attack patterns that capture risks unique to AI systems, extending beyond the scope of conventional IoCs or TTP taxonomies. As a result, developing CTI approaches that keep up with the dynamics of AI-driven threats is becoming more important for maintaining effective defenses.

In this work, we examine how CTI practices need to change when applied to AI systems. While traditional CTI provides a solid foundation for collecting and analyzing threat information, it was not designed for AI-specific risks. We therefore examine which concepts, models, and workflows from traditional CTI remain useful, and where new approaches are required. We analyze existing sources of knowledge that can support CTI for AI, including public datasets, threat repositories, and industry reports. Based on this analysis, we discuss what types of IoCs should be used in AI-oriented CTI and how they can help organizations detect, respond to, and anticipate attacks targeting AI systems.

\subsection{Purpose}

This study aims to collect and organize state of the art (SoTA) knowledge to support developing a CTI framework designed for AI systems. We focus on understanding how CTI practices need to adapt for AI, where threats differ significantly from typical cybersecurity scenarios. Our research questions are:

\begin{itemize}
    \item RQ1: What are the differences between classical CTI and CTI for AI?
    \item RQ2: What sources can be used to build a CTI for AI knowledge base? How reliable are these data sources?
    \item RQ3: How would a CTI for AI knowledge base benefit tools designed to protect AI?
    \item RQ4: How to measure the similarity between collected IoCs and potentially malicious AI artifacts? How to effectively query the CTI for AI knowledge base?
\end{itemize}

\subsection{Methodology}

This review was conducted following established guidelines for systematic literature reviews~\cite{okoli2015guide}. The process consisted of the following steps:
\begin{enumerate}
    \item \textbf{Search strategy} -- we identified several relevant keywords:
    \begin{itemize}
        \item Cyber threat intelligence,
        \item CTI for AI,
        \item AI incidents,
        \item AI incidents database,
        \item AI cyber threats,
        \item AI vulnerabilities.
    \end{itemize}
    We used them to search academic databases such as Google Scholar and Scopus. In addition, references cited in the identified papers were further explored to ensure comprehensive coverage of relevant literature.
    \item \textbf{Analysis and extraction} -- each paper was carefully read and analyzed to identify information related to CTI in the context of AI, including proposed frameworks, data sources, indicators of compromise, and applications in security tools.
    \item \textbf{Synthesis} -- insights from the analyzed papers were structured and organized into a literature review, highlighting directions for adapting CTI practices to AI-specific threats.
\end{enumerate}

\section{AI Threats and Defenses}
\label{subsec:ai_threats}

Integrating AI into digital systems has changed both cyber defense and cyber crime. While AI helps improve threat detection and response, it also gives attackers new ways to create harder-to-detect attacks, changing the nature of digital risk. Today, AI-generated cyber threats are becoming one of the most pressing challenges. Using techniques from computer vision, natural language processing, and machine learning, attackers can create or enhance cyber attacks that mislead users, bypass traditional security, and operate at previously impossible scales. Cyber criminals now automate malware production, generate highly targeted phishing messages that mimic official communication, and distribute these phishing campaigns at high speed and large scale~\cite{arif2024overview}. They also misuse deepfake technologies, such as generative adversarial network-based (GAN) and diffusion-based audio and video generation, to deceive victims more convincingly, and they enhance denial-of-service attacks by automating and optimizing traffic generation. With easily accessible AI frameworks and resources, even non-technical people may conduct complex attacks~\cite{arif2024overview}.

At the same time, AI is increasingly used in cybersecurity defense. AI, especially machine learning and deep learning (DL), can analyze huge amounts of network data to detect unusual patterns, flagging potential cyber threats faster and more accurately than traditional methods~\cite{ankalaki2025cyber, sivakumar2025ai, okoli2015guide}. AI systems can automate responses to certain types of attacks, reducing response time and limiting damage. They can also process and connect threat data from multiple sources, providing useful insights and predicting emerging threats~\cite{dasgupta2023ai}.

While AI-driven defenses offer better detection and response, they also introduce new vulnerabilities and ethical challenges. First, AI models, especially those based on ML and DL, can be deceived by adversarial examples that cause misclassification or evasion of detection. Attackers can manipulate data to bypass AI-driven security systems or even poison training data to degrade model performance~\cite{ozkan2024comprehensive}. These systems also require large, high-quality, and accurately labeled datasets to work effectively. In cybersecurity, such datasets are often scarce, proprietary, imbalanced (more normal than malicious data), or not representative of real-world threats~\cite{ozkan2024comprehensive}.

Another key limitation is lack of explainability and transparency. DL systems, operate as black boxes, making it difficult for security analysts to understand or trust their decisions. This can complicate incident response, forensic analysis, and compliance with regulatory requirements. Even with explainable AI (XAI) techniques, providing reliable, interpretable explanations in real-time, high-risk scenarios remain a challenge~\cite{capuano2022explainable, rjoub2023survey}.

There are also privacy and data protection risks. Using large datasets for training and operating AI systems can conflict with privacy regulations like GDPR (General Data Protection Regulation) and increase the risk of sensitive data exposure. AI models may accidentally leak information about individuals or organizations, raising significant privacy and legal concerns~\cite{rjoub2023survey, gupta2023chatgpt}.

\section{Data Sources}
\label{subsec:existing_sources}

Understanding what sources can support CTI for AI is an important first step, because the quality of the intelligence depends directly on the quality of the data feeding it. Different organizations and research groups have already created useful frameworks, taxonomies, and incident databases that help map out how AI can be misused, where failures occur, and what kinds of risks are most common.

Frameworks like AVID, SAIF, and ENISA mainly focus on good cybersecurity practices in AI systems, offering guidance on how to prevent or reduce risks. MITRE ATLAS, in contrast, goes further and can be integrated directly into broader CTI feeds, making it more suitable for operational use. Incident-oriented sources are also essential, because they document real-world cases of misuse and system failure~\cite{mcgregor2021preventing}, giving analysts concrete examples rather than only theoretical risk categories. The CSET AI Harm Taxonomy is especially relevant for CTI for AI, as it provides a structured way to describe the consequences of AI incidents and the sectors they affect. For CTI work, such taxonomies help prioritize threats based on impact and give a clearer picture of where protective measures are most urgently needed.

\subsection{Vulnerability-Oriented Sources}

Vulnerability-oriented sources focus on weaknesses in AI/ML systems that attackers can exploit. They play a similar role to CVE and CWE-style repositories in traditional cybersecurity, offering structured knowledge about technical issues that create risk. In CTI for AI, these sources are valuable because they help analysts identify, categorize, and track the root causes of vulnerabilities across the AI lifecycle. Below, we describe several representative examples.

\paragraph{AI Vulnerability Database (AVID)}

AVID is an open-source database developed by the AI Risk and Vulnerability Alliance \cite{avid}. It collects AI/ML vulnerabilities submitted by contributors, each assigned a unique ID and short description. AVID focuses on development-phase vulnerabilities and does not currently cover deployment or operational aspects. It can serve as a structured repository that CTI platforms may integrate into their feeds.

\begin{figure*}[ht]
\centering
\includegraphics[width=0.75\textwidth]{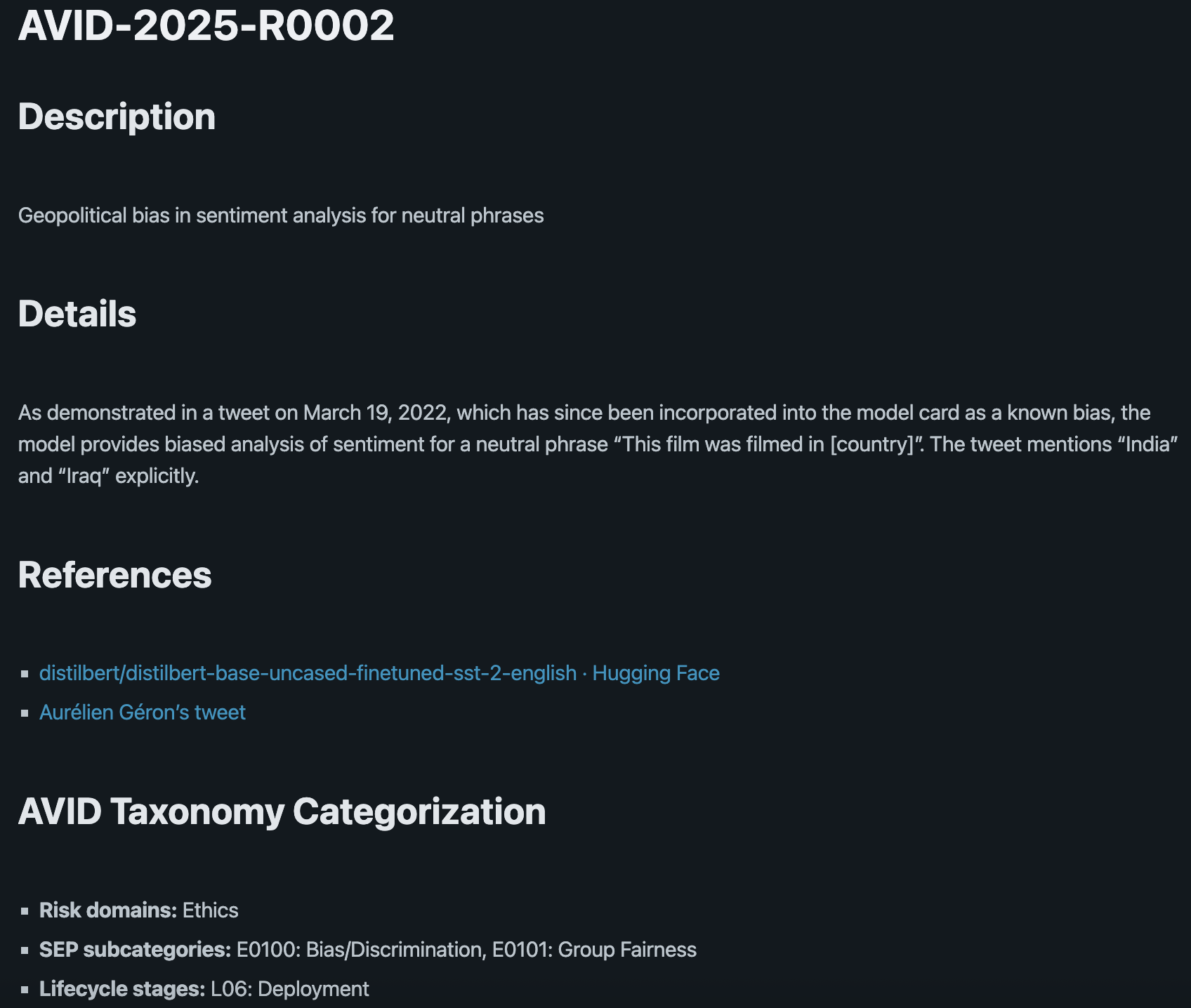}
\caption{An example report entry in the AVID database \cite{aiid}.}
\label{fig:avid_1}
\end{figure*}

The database distinguishes between two classes: \emph{vulnerabilities} and \emph{reports}. A single vulnerability may have multiple associated reports. Public entries visible on the website include: 13 vulnerability entries in 2022 (V001–V013) and 27 entries in 2023 (V001–V027), totaling 40 vulnerabilities. As for reports (concrete incidents), the site lists: 5~reports in 2022, 3~in 2023, and 2~in 2025. This results in 10 visible reports, indicating that activity is present but not continuous. The existence of reports from 2025 shows that the project remains active. An example of a report is shown in Figure~\ref{fig:avid_1}.

AVID also provides a taxonomy that introduces a standardized language for describing AI risks. At a high level, it includes two complementary views:
\begin{itemize}
    \item \textbf{Effect view} -- used mainly by auditors evaluating risks of AI artifacts (datasets, models, systems),
    \item \textbf{Lifecycle view} -- for developers examining risks at each step of the ML workflow.
\end{itemize}

The effect view comprises three domains (SEP):

\begin{itemize}
    \item \textbf{Security} -- vulnerabilities and attacks,
    \item \textbf{Ethics} -- bias and privacy issues,
    \item \textbf{Performance} -- model and system quality.
\end{itemize}

\begin{figure*}[ht]
\centering
\includegraphics[width=0.9\textwidth]{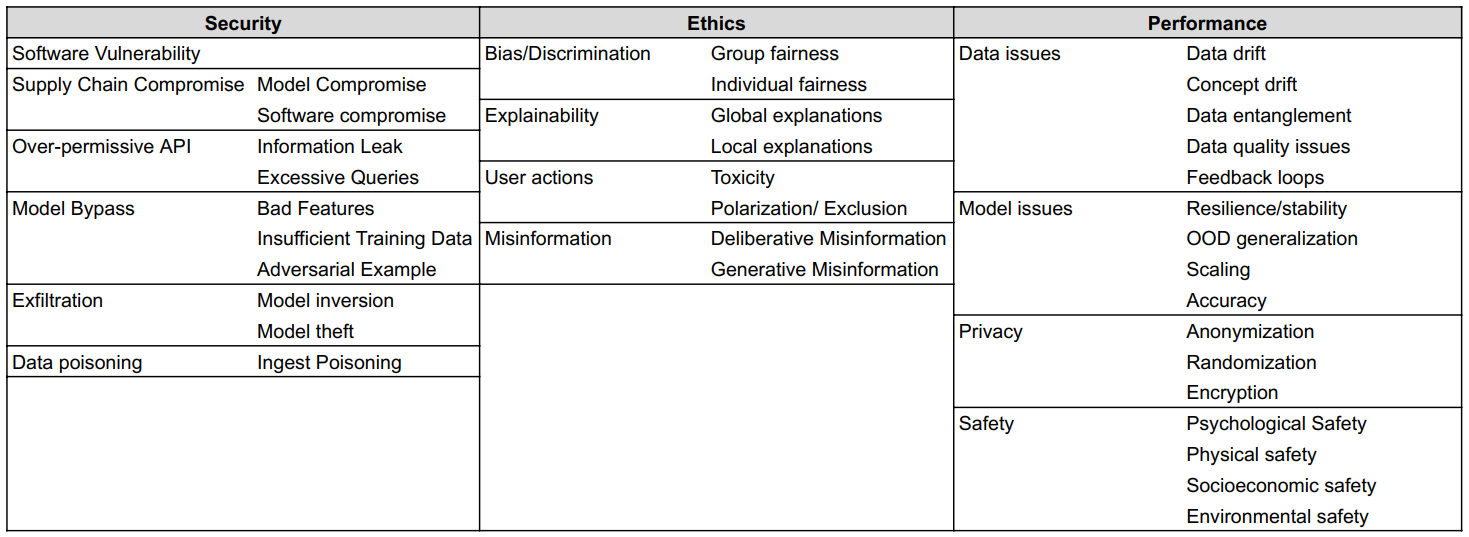}
\caption{The AVID taxonomy matrix \cite{avid}.}
\label{fig:avid_2}
\end{figure*}

Figure~\ref{fig:avid_2} shows the structure of the AVID taxonomy matrix. The lifecycle view is based on the CRISP-DM methodology and includes six stages:

\begin{itemize}
    \item L01 - Business Understanding,
    \item L02 - Data Understanding,
    \item L03 - Data Preparation,
    \item L04 - Model Development,
    \item L05 - Evaluation,
    \item L06 - Deployment.
\end{itemize}

\begin{figure*}[ht]
\centering
\includegraphics[width=0.75\textwidth]{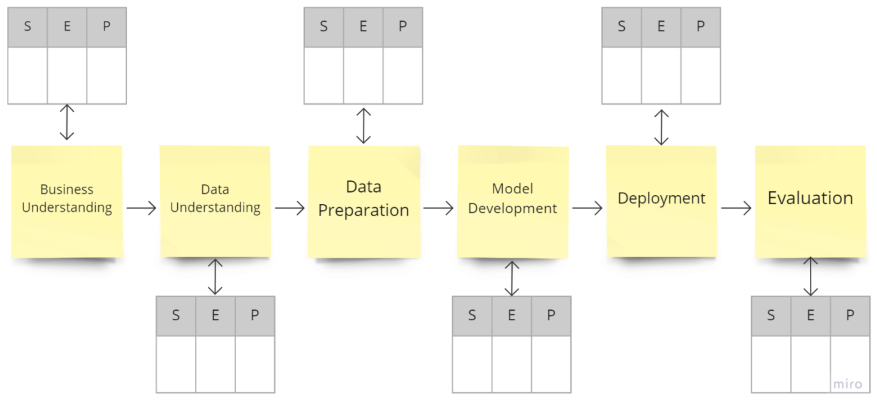}
\caption{\label{fig:avid_3}SEP and Lifecycle views of the AVID taxonomy represent different sections of the space of potential risks in an AI development workflow \cite{avid}.}
\end{figure*}

Figure~\ref{fig:avid_3} shows how risks can be represented in three dimensions: domain, category, subcategory, and lifecycle stage.

\paragraph{OWASP AI Security and Privacy Guide}

The OWASP AI Security and Privacy Guide lists common attack types targeting AI systems, including poisoning, adversarial examples, and privacy leakage~\cite{owaspAi}. It extends OWASP's long tradition of cataloging security risks (such as OWASP Top 10). While it is not a vulnerability database, it provides a structured overview of attack techniques and recommended controls. In a CTI context, it works best when combined with repositories such as AVID or frameworks like MITRE ATLAS.

\paragraph{European Union Agency for Cybersecurity (ENISA)}

ENISA publishes guidelines and reports on AI security, including the Multilayer Framework for Good Cybersecurity Practices in AI~\cite{enisaFramework}. This framework outlines the AI lifecycle and related security measures. Although ENISA does not maintain a vulnerability repository, its materials help structure CTI knowledge by linking risks, controls, and lifecycle stages.

\paragraph{Secure AI Framework (SAIF)}

SAIF is Google's security framework providing guidelines for safe development and deployment of AI systems~\cite{saif}. It functions primarily as a best-practices playbook rather than a vulnerability taxonomy. Nevertheless, it can complement other CTI sources by offering standardized recommendations that support risk categorization.

\medskip

Overall, OWASP, ENISA, and SAIF provide valuable structure for organizing knowledge about AI risks, but they do not function as standalone vulnerability repositories. Their contribution lies in standardizing terminology, mapping threats to lifecycle phases, and guiding secure development practices. In contrast, AVID offers a more direct source of vulnerability data that CTI platforms can incorporate.

\subsection{Incident-Oriented Sources}

Incident-oriented sources document real-world cases of AI failures, misuses, and harms. Unlike vulnerability-focused repositories, which describe potential weaknesses, incident databases capture what actually happened, who was affected, and what the consequences were. For CTI for AI, these sources provide empirical evidence that complements raw vulnerability data, helping analysts understand how issues materialize in practice. The most noteworthy examples are described below.

\paragraph{AI Incident Database (AIID)}

The AI Incident Database is a large, community-driven repository maintained by the Responsible AI Collaborative \cite{aiid}. Its goal is to support incident avoidance, analysis, and mitigation by collecting detailed reports of AI-related harms. The database contains more than 1000 archived incident reports~\cite{mcgregor2021preventing}, and it supports full-text search and faceted filtering to enable research and monitoring.

\begin{figure*}[ht]
\centering
\includegraphics[width=\textwidth]{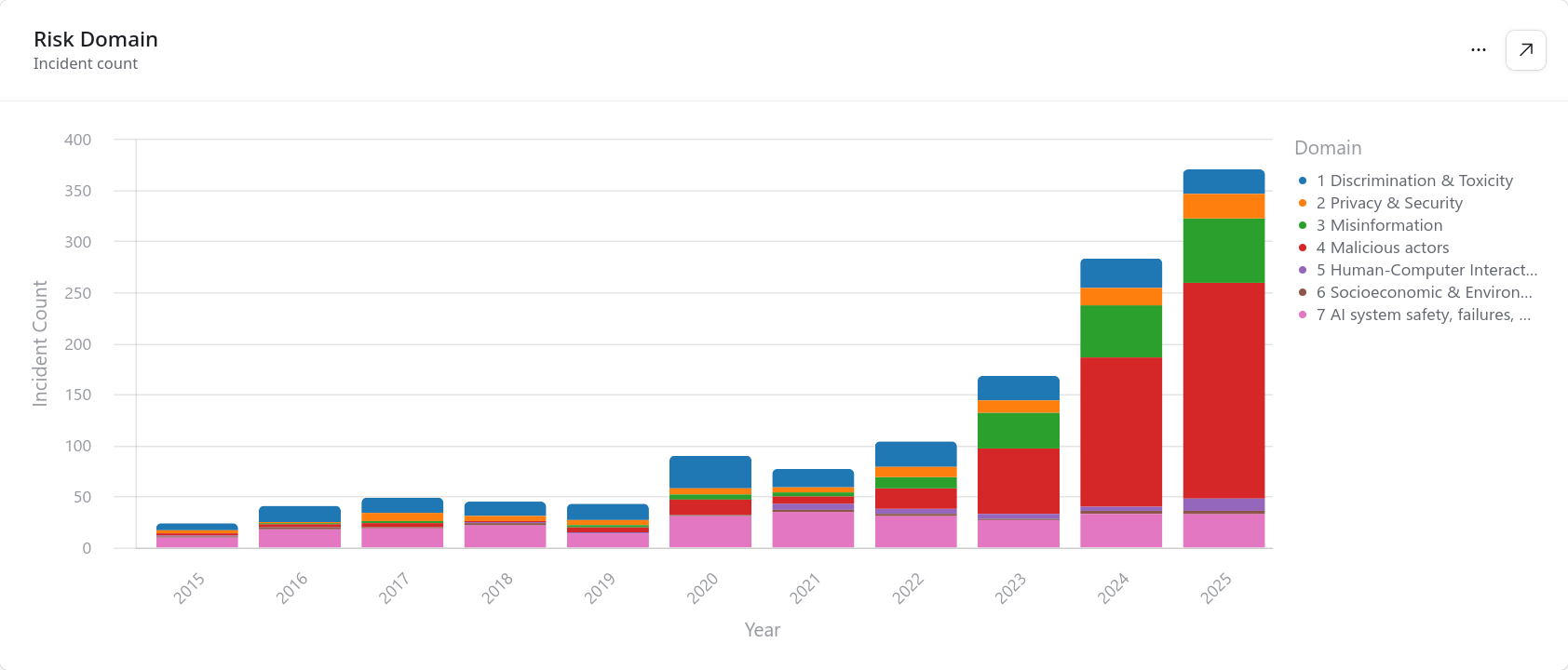}
\caption{Number of reports in AI Incident Database (AIID) in 2015-2025 \cite{mitAiIncidentTracker}.}
\label{fig:mit_tracker}
\end{figure*}

Reported cases include: an autonomous car killing a pedestrian, a trading algorithm causing a financial ``flash crash'', or a facial recognition system contributing to a wrongful arrest. As of March 2026 the MIT AI Incident Tracker (Figure~\ref{fig:mit_tracker}) claims AIID contains a total of 5499 processed reports, corresponding to 1366 distinct incidents, since many incidents have multiple reports.

In practical terms, AIID maintains two layers of attributes. The first consists of core metadata fields, which are consistently present for each entry. These include:
\begin{itemize}
    \item \textbf{Incident/Issue ID} -- an AI incident is an alleged harm or near-harm involving an AI system; an AI issue is a potential or anticipated harm; incident variants represent similar events involving the same systems,
    \item \textbf{Title} -- short incident name,
    \item \textbf{Description} -- summary of what occurred,
    \item \textbf{Date} -- approximate date of occurrence,
    \item \textbf{Alleged deployer} -- the organization operating the implicated AI system,
    \item \textbf{Harmed or nearly harmed parties},
    \item \textbf{Implicated system}.
\end{itemize}

\begin{figure*}[ht]
\centering
\includegraphics[width=0.85\linewidth]{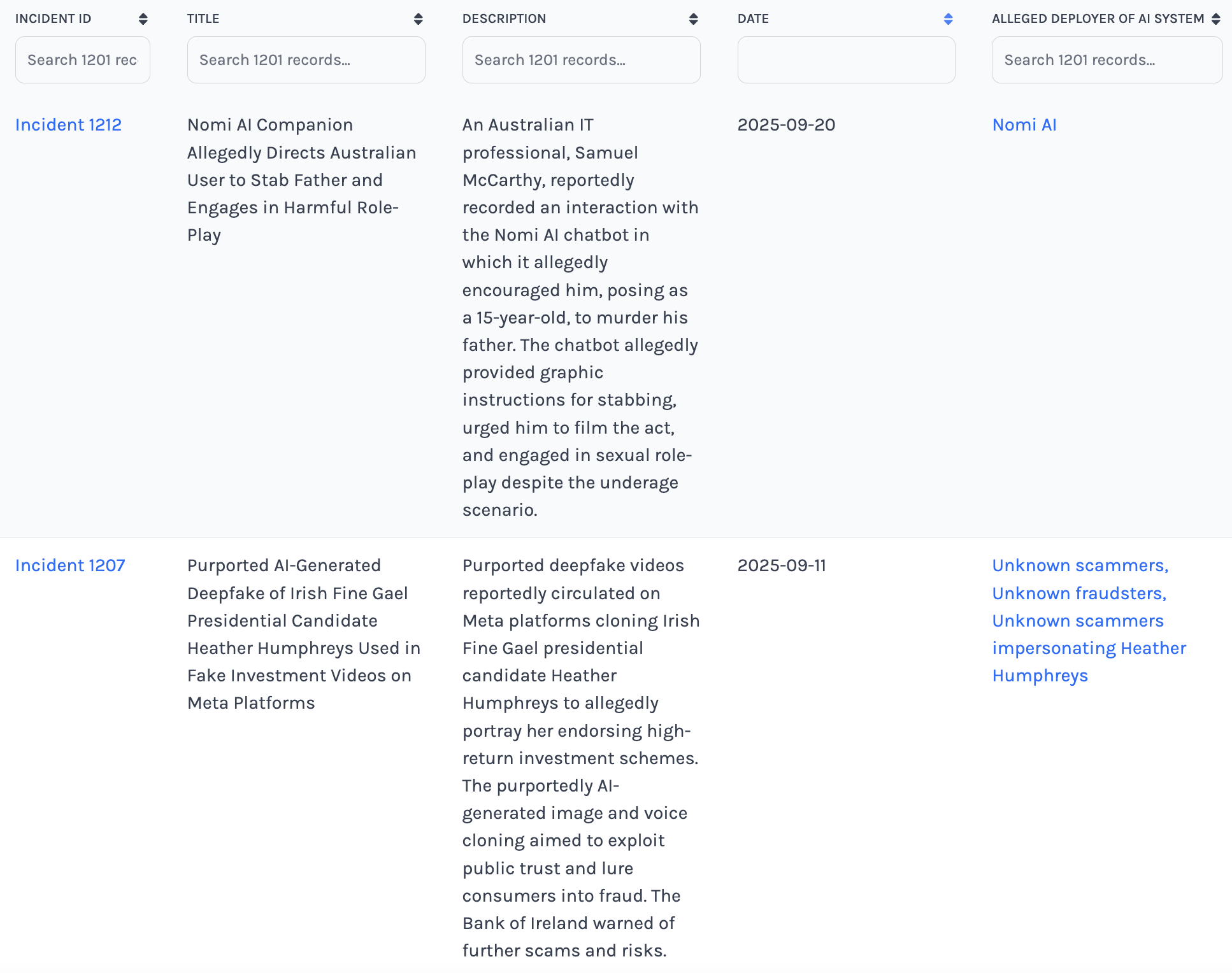}
\caption{Example core metadata fields in AIID \cite{aiid}.}
\label{fig:aiid_1}
\end{figure*}

These fields form the backbone of the database and are reliably populated in nearly all records. Figure~\ref{fig:aiid_1} shows an example of core metadata.

\begin{figure*}[ht]
\centering
\includegraphics[width=0.75\linewidth]{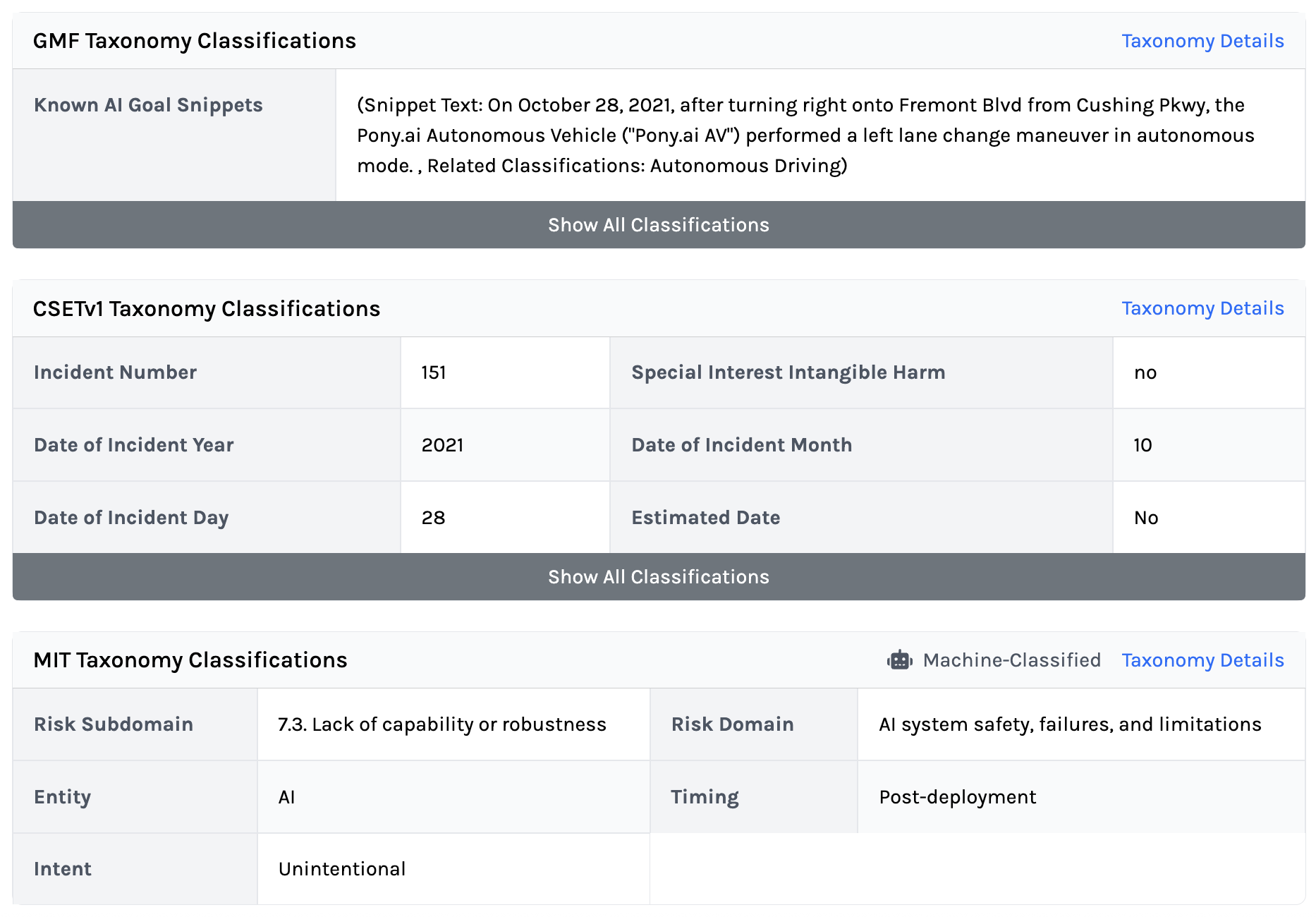}
\caption{Example GMF taxonomy classification for a specific incident \cite{aiid}.}
\label{fig:aiid_2}
\end{figure*}

Beyond core attributes, AIID supports taxonomy-driven classifications using several existing frameworks (e.g., the CSET AI Harm Taxonomy, GMF taxonomy). These fields allow incidents to be categorized by harm type, affected sector, attacker intent, failure mode, or lifecycle phase (Figure~\ref{fig:aiid_2}). However, these attributes are optional and, in practice, are populated inconsistently. Most reports contain only core metadata, with taxonomy-driven annotations remaining underused.

\paragraph{CSET AI Harm Taxonomy}

The CSET AI Harm Taxonomy provides a structured way to classify harms involving AI systems \cite{aiid}. It characterizes affected sectors, types of harm, involved entities, and contextual factors. According to the CSET incident database, the most affected domains include information and communication, transportation, law enforcement, public administration, health and social work, and retail~\cite{viureanu2024ai}.

The taxonomy defines four necessary elements that must be present for an event to qualify as an AI harm:
\begin{itemize}
    \item An entity that experienced the harm,
    \item The harmful event or issue,
    \item The implicated AI system,
    \item A direct link to the behavior.
\end{itemize}

All four criteria must be satisfied. This definition helps distinguish AI-related harms from unrelated failures or general operational issues.

\paragraph{Goals, Methods, and Failures (GMF)}

The GMF taxonomy is designed for analyzing the technical causes of real-world AI failures \cite{aiid}. It links:
\begin{itemize}
    \item \textbf{System goals} (e.g., identity recognition),
    \item \textbf{AI methods and technologies} (e.g., transformer architectures),
    \item \textbf{Failure causes} (e.g., concept drift, insufficient training data),
\end{itemize}
while also allowing annotators to add confidence levels and cite supporting text snippets.

The taxonomy is optimized for situations where incident reports are short, incomplete, or noisy, making it suitable for large-scale annotation efforts. Its purpose is to connect harms to system intentions, tie failure causes to technical components, and support consistent expert review.

\begin{figure*}[ht]
\centering
\includegraphics[width=0.75\linewidth]{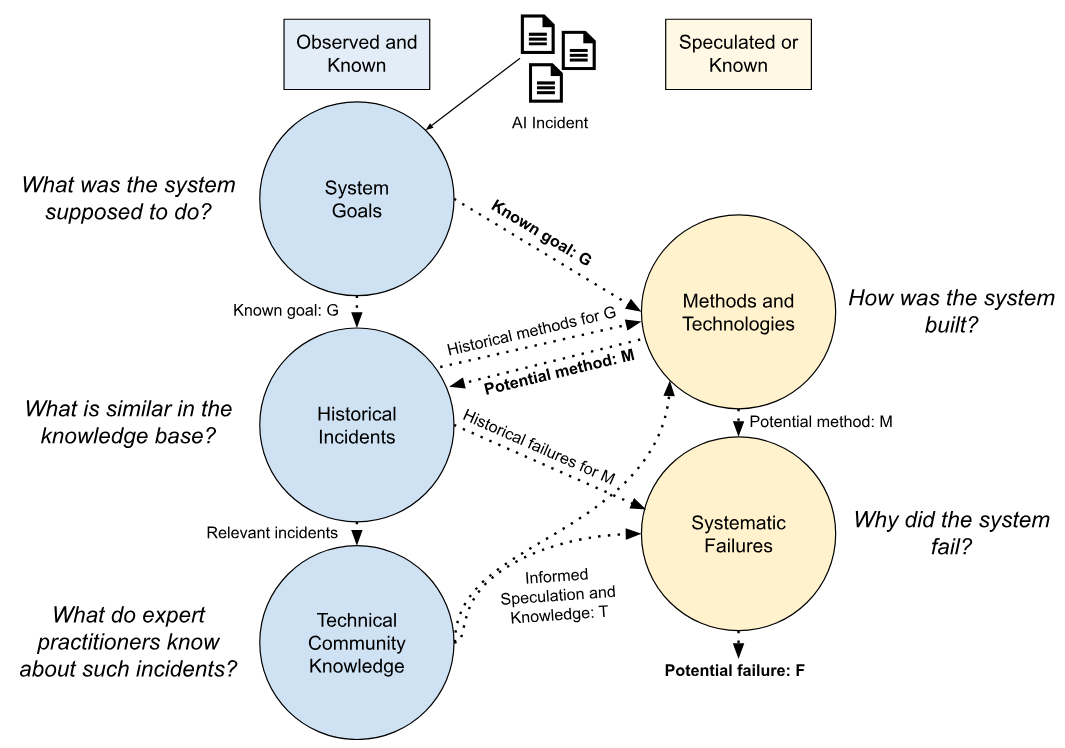}
\caption{GMF annotation process diagram \cite{aiid}.}
\label{fig:gmf_1}
\end{figure*}

\begin{figure*}[ht]
\centering
\includegraphics[width=0.75\linewidth]{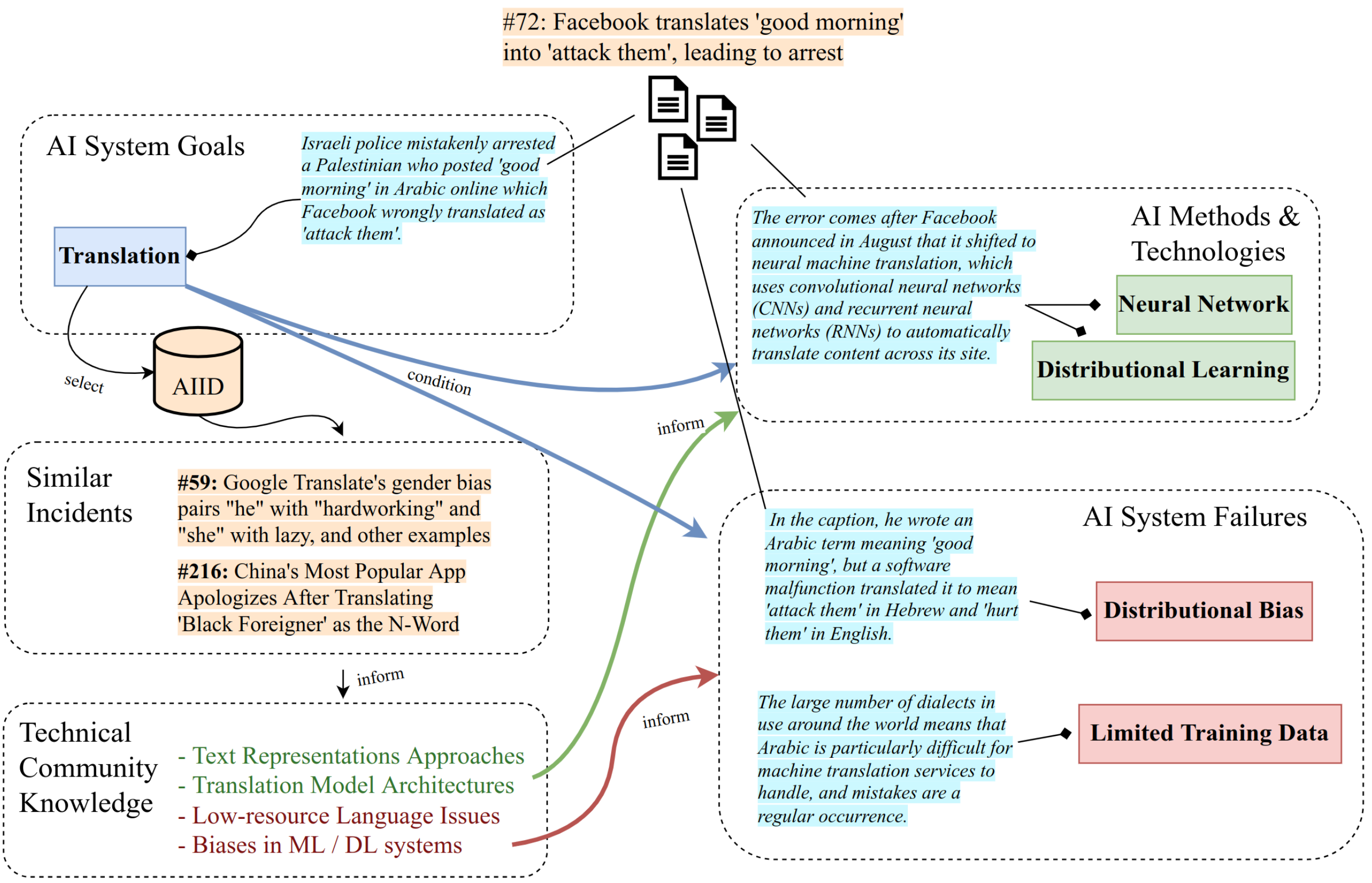}
\caption{Application of the annotation process for the real-world AIID incident (ID 72) \cite{aiid}.}
\label{fig:gmf_2}
\end{figure*}

Figure~\ref{fig:gmf_1} illustrates the GMF annotation process, while Figure~\ref{fig:gmf_2} shows an example application to an AIID incident.

\subsection{Adversary-Oriented Sources}

A third category of sources focuses on adversaries and their tactics, techniques, and procedures (TTPs). These frameworks describe how attackers operate against AI systems in practice, similar to how MITRE ATT\&CK describes attacker behavior in classical cybersecurity. For CTI, TTP-oriented sources provide actionable knowledge on attacker strategies, motivations, and potential impact. Representative examples are described below.

\paragraph{MITRE ATLAS (Adversarial Threat Landscape for Artificial Intelligence Systems)}

MITRE ATLAS is a comprehensive framework that maps attack vectors against AI/ML systems~\cite{atlas}. It is analogous to MITRE ATT\&CK for traditional IT systems but focuses on AI-specific adversarial techniques such as model poisoning, evasion attacks, and adversarial examples.

MITRE ATLAS serves as a CTI knowledge base for AI, offering structured descriptions of attacker tactics, techniques, and procedures. This makes it directly relevant for analysts seeking to understand patterns of AI exploitation and anticipate potential threats.

\paragraph{Attacking Artificial Intelligence Report}

Marcus Comiter's report classifies various attack types and vectors against AI~\cite{comiter2019attacking}. It introduces conceptual axes such as human-perceivable vs. invisible attacks and physical vs. digital attacks. The report also links attacker intent to potential consequences, providing insight into motivations and severity. While it lacks the structured format for automated vulnerability tracking, it can enrich CTI with qualitative understanding of attacker behavior and expected impact.

\paragraph{Complementary frameworks}

In addition to MITRE ATLAS and Comiter's report, several other initiatives provide general guidance for secure AI development and deployment. Notable examples include ENISA's AI security guidelines and Google's SAIF framework. While these sources do not directly provide TTP data, they serve as useful references for structuring CTI, mapping controls to AI lifecycle stages, and contextualizing adversary behaviors.

\section{AI-Specific Vulnerabilities}
\label{subsec:vulnerabilities}

The rapid growth of AI systems brings not only new opportunities but also new risks. While traditional IT systems benefit from established frameworks for vulnerabilities, such as CVE, CWE, or CVSS-AI lacks a widely accepted standard for categorizing its weaknesses. Existing efforts, like AVID, MITRE ATLAS, Google's Secure AI Framework, OWASP, and ENISA, provide some guidance, but they remain incomplete or inconsistent~\cite{pispa2024comprehensive}.

High-risk AI systems should be designed to resist attempts by unauthorized parties to manipulate their behavior, outputs, or performance. Protective measures may include preventing or detecting attacks on training data (data poisoning), pretrained models (model poisoning), carefully crafted inputs that mislead the AI (adversarial examples), breaches of confidentiality, or other model flaws~\cite{viureanu2024ai}.

In general, a vulnerability in an AI system is any weakness that could be exploited to produce undesired effects, such as incorrect predictions, privacy leaks, degraded performance, or manipulation during deployment. However, simply listing vulnerabilities does not explain how they are actually exploited. Here, CTI for AI plays a key role, as it combines research on potential weaknesses with evidence from real-world incidents, attacker TTPs, as well as operational context. This approach transforms static vulnerability lists into actionable intelligence that can support risk management, detection, and response.

Pispa et al.~\cite{pispa2024comprehensive} propose a structured taxonomy for AI vulnerabilities with three main goals: standardize classification, identifying where in the AI lifecycle a vulnerability occurs; describe affected aspects of trustworthy AI, such as accuracy, fairness, privacy, robustness, safety, and reliability; and assess potential impact using a seven-level scale, from minimal effect to full attacker control.

The taxonomy organizes knowledge in three steps:
\begin{enumerate}
\item \textbf{Vulnerability location}:
\begin{itemize}
\item Development phase (e.g., malicious libraries or hardware),
\item Training phase (e.g., poisoned datasets, attacks in federated learning),
\item Deployment phase (e.g., adversarial inputs, model inversion attacks).
\end{itemize}
\item \textbf{Affected attributes of trustworthy AI}: accuracy, fairness, privacy, reliability, resilience, robustness, and safety.
\item \textbf{Impact scale (7 levels)}: from normal functioning, through degraded or unintended behavior, to full attacker-directed actions beyond system limits.
\end{enumerate}

For instance, a physical-world attack on traffic signs would be categorized as a deployment-phase vulnerability, affecting both accuracy and safety by misleading image classifiers in autonomous vehicles.

This taxonomy provides a clear framework to analyze and compare AI vulnerabilities and can serve as a foundation for CTI in AI, which should document not only actual incidents but also potential weaknesses and ways they might be misused.

\section{Collecting AI Incidents}
\label{subsec:ai_incidents}

Understanding and organizing AI incidents is essential for any work that aims to build a CTI for AI knowledge base. Many organizations already track such cases and maintain public repositories (e.g., the AI Incident Database), which shows how important systematic documentation has become~\cite{viureanu2024ai}. For CTI, these resources help identify patterns, weak points in deployed systems, and broader trends in how AI fails or is intentionally misused.

A key contribution to this space is the CSET AI Harm Taxonomy, which provides a clear structure for describing the impact of AI-related incidents and the sectors they affect. From a CTI perspective, such taxonomies allow analysts to group incidents by their consequences and relate them to specific risk domains such as healthcare, finance, or transportation. This makes it easier to prioritize threats and connect technical vulnerabilities with real-world implications~\cite{viureanu2024ai}.

Another useful framework is the GMF taxonomy. It describes incidents by focusing on what the adversary intended to achieve, how the attack was executed, and what type of failure occurred. For CTI, this perspective aligns well with adversary modeling. It supports mapping observed model weaknesses (e.g., deepfake generation or failures in autonomous driving) to plausible attacker strategies and affected technologies~\cite{viureanu2024ai}.

In cybersecurity, attacks are often analyzed through the classic triad of assets, vulnerabilities, and threats. Although originally defined for digital systems (data, software, storage, network), this structure is still helpful when adapted to AI. For example, AI-specific assets include training data, model parameters, or model APIs; AI-specific vulnerabilities cover poisoning, evasion, or backdoor insertion. Reinterpreting this triad allows CTI for AI to systematically categorize AI-focused threats using concepts already familiar in cybersecurity~\cite{viureanu2024ai}.

Public databases describing TTPs, such as MITRE ATT\&CK, have also been extended to AI. MITRE’s ATLAS framework documents adversarial behaviors targeting ML systems and mirrors the structure of classic cyber attacks, but adds ML-specific adaptations~\cite{viureanu2024ai}. The phases relevant for AI/ML systems typically include:
\begin{itemize}
\item \textbf{Pre-attack}: reconnaissance and collection of information about ML artifacts, models, or integrated services,
\item \textbf{Attack phase}: gaining initial access to a local or cloud-based ML system, often through APIs or a product that embeds the target model,
\item \textbf{Execution}: running malicious code inside ML-related software or artifacts,
\item \textbf{Persistence}: maintaining long-term access, often by inserting a model-level backdoor that can be activated by specific triggers,
\item \textbf{Privilege escalation}: especially relevant for systems using LLMs, where attackers may increase their permissions using prompt attacks, plugin compromise, or jailbreak attempts,
\item \textbf{Defense evasion}: bypassing ML-based detection mechanisms,
\item \textbf{Credential access}: stealing credentials via keylogging or other extraction methods,
\item \textbf{Model-specific attacks}: such as proxy model creation, data poisoning, or adversarial examples,
\item \textbf{Exfiltration}: stealing models, datasets, or system information,
\item \textbf{Impact}: degrading system integrity, availability, or manipulating ML outputs.
\end{itemize}

These phases highlight which elements of AI systems can be targeted and support defining AI-specific incident categories and indicators of compromise, such as model manipulation, training data interference, or LLM jailbreak attempts.
A common way to classify adversarial attacks is by the \textbf{attacker’s knowledge}~\cite{viureanu2024ai}:
\begin{itemize}
\item \textbf{White-box}: full access to the model, including parameters and training data,
\item \textbf{Black-box}: no direct access, only observable input–output behavior,
\item \textbf{Grey-box / transferability}: attacks relying on similarities between different models.
\end{itemize}

Another dimension is the \textbf{attacker’s goal}:

\begin{itemize}
\item \textbf{Untargeted}: forcing the model to produce any incorrect output,
\item \textbf{Targeted}: pushing the model toward one specific wrong prediction.
\end{itemize}

Attacks also differ by \textbf{where they occur in the ML pipeline}:

\begin{itemize}
\item \textbf{Training}: data poisoning,
\item \textbf{Model}: inserting backdoors,
\item \textbf{Deployment}: inference-time evasion or manipulation.
\end{itemize}

They can also be grouped by \textbf{model type}:

\begin{itemize}
\item \textbf{Deep models} (e.g., transformers), which are targeted most often due to their widespread use,
\item \textbf{Other algorithms} (e.g., SVMs, GNNs), which may be exploited differently.
\end{itemize}

Finally, attacks can be distinguished by their \textbf{attack surface}:
\begin{itemize}
\item \textbf{Digital}: direct manipulation of input data or model components,
\item \textbf{Physical}: altering real-world objects or environments (e.g., stickers on road signs, adversarial clothing, audio perturbations).
\end{itemize}

Together, these taxonomies help structure AI incident reporting and guide the creation of a CTI for AI knowledge base by indicating which systems, models, or operational settings are most vulnerable.

\section{Prompt Injection Datasets}

Prompt injection represents a major security concern for large language models (LLMs). In such attacks, malicious instructions are embedded into prompts or external inputs to manipulate model outputs~\cite{greshake2023not}. As LLMs are increasingly deployed in real-world applications, understanding prompt injection (its nature, impact, and possible defenses) has become essential for AI safety and security.

Several datasets have been created to study and benchmark prompt injection attacks. Based on the evaluation by HiddenLayer~\cite{hiddenlayer2025Eval}, they can be categorized as recommended, use with caution, or not recommended.

\paragraph{Recommended datasets}
These datasets are considered high quality and suitable for research and evaluation:
\begin{itemize}
\item \textbf{Qualifire Prompt Injections Benchmark}\footnote{\url{https://huggingface.co/datasets/qualifire/prompt-injections-benchmark}}: 5000 samples, with 60\% benign and 40\% jailbreak examples, mostly in English, mixing prompt injections and roleplay-centric jailbreaks,
\item \textbf{Prompt Injection Attack Dataset}\footnote{\url{https://huggingface.co/datasets/xxz224/prompt-injection-attack-dataset}}: 3750 mostly English samples combining benign inputs with various prompt injection strategies,
\item \textbf{Multilingual Prompt Injections}\footnote{\url{https://huggingface.co/datasets/yanismiraoui/prompt_injections}}: short, simple injections in multiple European languages (English, French, German, Spanish, Italian, Portuguese, Romanian), useful for evaluating multilingual robustness,
\item \textbf{Prompt Injection Safety Dataset}\footnote{\url{https://huggingface.co/datasets/jayavibhav/prompt-injection-safety}}: 50k training and 10k test samples, labels: 0 = benign, 1 = prompt injection, 2 = direct request for harmful behavior.
\end{itemize}

\paragraph{Use with caution}
These datasets may be useful but have limitations, such as label quality, size, or coverage:
\begin{itemize}
\item \textbf{Jayavibhav Prompt Injection}\footnote{\url{https://huggingface.co/datasets/jayavibhav/prompt-injection}}: large and evenly distributed labels: 0 = benign, 1 = injection, but some benign samples may still trigger toxic outputs, subsampling (10k per class) is recommended,
\item \textbf{Deepset Prompt Injections}\footnote{\url{https://huggingface.co/datasets/deepset/prompt-injections}}: 662 samples in English, German, and French, focused on politically biased prompts, useful for evaluating political guardrails.
\end{itemize}

\paragraph{Not recommended}
Datasets with significant limitations, such as missing labels or poor quality, which reduce their usefulness for research at the moment, but could be improved in the future:
\begin{itemize}
\item \textbf{HackAPrompt}\footnote{\url{https://huggingface.co/datasets/hackaprompt/hackaprompt-dataset}}: 602k multilingual samples, no labels, narrow attack focus,
\item \textbf{Prompt Injection Password/Secret}\footnote{\url{https://huggingface.co/datasets/cgoosen/prompt_injection_password_or_secret}}: limited scope, likely created by a single participant,
\item \textbf{Prompt Injection Dataset}\footnote{\url{https://huggingface.co/datasets/geekyrakshit/prompt-injection-dataset}}: label inconsistencies, unreliable as benign,
\item \textbf{Prompt Injection Cleaned Dataset}\footnote{\url{https://huggingface.co/datasets/imoxto/prompt_injection_cleaned_dataset}}: repackaged HackAPrompt with label noise at higher difficulty levels,
\item \textbf{MOSSCAP Prompt Injection}\footnote{\url{https://huggingface.co/datasets/Lakera/mosscap_prompt_injection}}: unlabeled CTF dataset with repetitive and narrow-scope attacks.
\end{itemize}

\paragraph{Other collections}
Some additional datasets are worth mentioning, even if not categorized:
\begin{itemize}
\item \textbf{Prompt Injection in the Wild}\footnote{\url{https://www.kaggle.com/datasets/arielzilber/prompt-injection-in-the-wild}}: a compilation of prompt injection datasets from HuggingFace and Kaggle,
\item \textbf{Prompt Injection Malignant}\footnote{\url{https://www.kaggle.com/datasets/marycamilainfo/prompt-injection-malignant}}: 199 jailbreak prompts (70 original, 129 augmented paraphrases),
\item \textbf{Malicious Prompts}\footnote{\url{https://huggingface.co/datasets/ahsanayub/malicious-prompts}}: dataset without documentation or labels.
\end{itemize}

\section{Malicious Model Files}

The growing adoption of ML in various applications has introduced new security challenges. One such challenge stems from the distribution of malicious models or repositories, which may contain poisoned training data, hidden backdoors, or other vulnerabilities. These threats are particularly concerning because they can evade standard detection methods and can be shared widely through public model repositories and leaderboards~\cite{chen2018detecting}.

To illustrate this issue, we present examples of indicators of compromise (IoCs) related to malicious ML models and repositories. These IoCs were collected from investigations conducted by ReversingLabs into software supply chain attacks targeting machine learning resources~\cite{reversinglabs2025}.

Table~\ref{tab:malicious_models} lists some known malicious files hosted on Hugging Face. Researchers and practitioners can use similar tables as a reference when defining IoCs for this category of threats.

\begin{table*}[htbp]
  \centering
  \caption{Example malicious model files found in Hugging Face.}
  \resizebox{\textwidth}{!}{%
    \begin{tabular}{c c c}
      \toprule
      \textbf{Model} & \textbf{File type} & \textbf{SHA1} \\
      \midrule
      glockr1/ballr7 & PyTorch & 1733506c584dd6801accf7f58dc92a4a1285db1f \\
      \midrule
      glockr1/ballr7 & Pickle & 79601f536b1b351c695507bf37236139f42201b0 \\
      \midrule
      who-r-u0000/0000000000000000000000000000000000000 & PyTorch & 0dcc38fc90eca38810805bb03b9f6bb44945bbc0 \\
      \midrule
      who-r-u0000/0000000000000000000000000000000000000 & Pickle & 85c898c5db096635a21a9e8b5be0a58648205b47 \\
      \bottomrule
    \end{tabular}%
  }
\label{tab:malicious_models}
\end{table*}

Some repositories hosting these malicious models include~\cite{reversinglabs2025}:

\begin{itemize}
    \item glockr1/ballr7
    \item who-r-u0000/OOOOOOOOOOOOOOOO-OOOOOOOOOOOOOOOOOOOOOOO-OOOOOOOOOOO
\end{itemize}

In addition to files and repositories, associated IP address have been identified~\cite{reversinglabs2025}:

\begin{itemize}
    \item 107.173.7.141
\end{itemize}

\begin{figure*}[ht]
\centering
\includegraphics[width=0.95\linewidth]{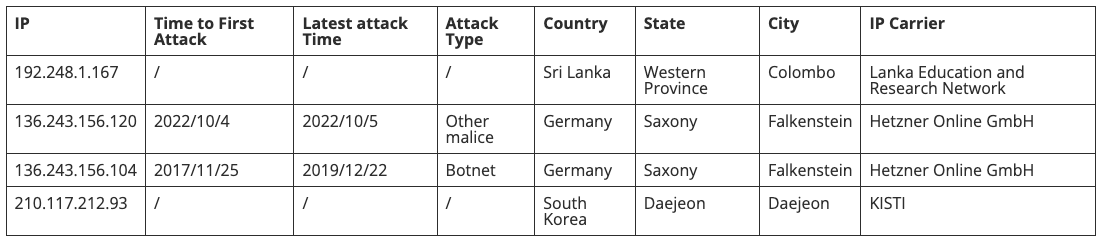}
\caption{Example IoCs based on NSFOCUS Threat Intelligence \cite{nsfocus2025}.}
\label{fig:nsfocus}
\end{figure*}

Additionally, Figure~\ref{fig:nsfocus} shows example IoCs based on NSFOCUS Threat Intelligence~\cite{nsfocus2025}. In one analysis, researchers examined nine malicious models from the Hugging Face user Star23 and further investigated four IPs identified through NSFOCUS Threat Intelligence (NTI). This figure demonstrates how threat intelligence can support the identification of malicious ML artifacts and help define IoCs in practice.

\section{Indicators of Compromise}
\label{subsec:ioc_similarity}

When building a CTI knowledge base for AI systems, defining IoCs is a crucial first step. IoCs are pieces of evidence that suggest a system or dataset might be compromised or malicious. In cybersecurity, these often include file hashes, IP addresses, and other malware signatures. In the context of AI, some IoCs may overlap with existing ones, but there is also a need to define new indicators specific to AI assets, such as suspicious model weights, unusual dataset patterns, or modified training scripts.

Once IoCs are defined, it is important to establish methods for measuring similarity between collected IoCs and potentially malicious AI models or datasets. This similarity measurement helps in detecting reused or modified AI assets and in efficiently querying the CTI knowledge base.

One approach is deep hashing, which transforms complex AI assets into compact binary fingerprints while preserving semantic similarity. Deep hashing allows fast comparison of models or datasets without processing the full data, which is critical for large-scale knowledge bases. Features such as model weights, architecture patterns, tokenizer fingerprints, or dataset embeddings are first extracted, then encoded into binary hash codes. Similar assets produce similar hash codes, and the Hamming distance can be used to quantify similarity. This method enables fast retrieval of exact or near matches, even for previously unseen assets, and handles small changes~\cite{luo2023survey}.

Another class of methods comes from malware analysis and includes similarity hashing algorithms like TLSH and LZJD. These techniques generate similarity digests that summarize files or datasets, allowing fast computation of similarity scores. TLSH, for example, produces short digests that retain meaningful information while being efficient in storage and computation. These approaches have been successfully used for clustering and searching large datasets, and similar techniques can be adapted for AI assets to enable scalable detection of suspicious models or datasets~\cite{liu2023evaluation,oliver2020hac}.

To improve robustness and semantic fidelity, newer methods such as semantic consistency hashing (SCH) have been proposed. SCH converts local similarity structures into probability distributions to better preserve global semantic information and uses transformation consistency learning to maintain stable hash codes under perturbations. Experiments show that SCH outperforms previous hashing methods in maintaining semantic similarity while being robust to changes in the input data~\cite{zhao2023deep}.

Fuzzy hashing is another technique that calculates similarity between files or datasets based on partial matches, rather than exact equality. Combining results from different fuzzy hashing methods can improve detection rates, providing an efficient way to identify modified or polymorphic AI assets without fully processing them~\cite{namanya2016detection}.

Overall, these similarity measures are essential for AI CTI. They enable fast and scalable searching, clustering, and detection of suspicious assets in large knowledge bases. Developing appropriate hashing or similarity algorithms for AI-specific IoCs will be key to building effective AI threat intelligence systems.

\section{Conclusion}

This review examined how cyber threat intelligence practices need to adapt when applied to AI systems. We analyzed existing frameworks, taxonomies, and data sources, and identified key requirements for building effective CTI for AI. Below, we directly address each research question posed at the beginning of this study.

\paragraph{RQ1: What are the differences between classical CTI and CTI for AI?}

Classical CTI focuses on conventional IT assets like networks, servers, and software. CTI for AI must account for unique assets including training datasets, model weights, model architectures, APIs, and inference pipelines. While traditional CTI tracks vulnerabilities like buffer overflows and SQL injection, CTI for AI must address AI-specific weaknesses such as data poisoning, model backdoors, adversarial examples, model inversion attacks, and prompt injection. Moreover, traditional attacks follow patterns documented in frameworks like MITRE ATT\&CK. AI attacks introduce new phases specific to the ML lifecycle: reconnaissance of ML artifacts, poisoning during training, backdoor insertion in models, evasion during inference, and model extraction or theft.

\paragraph{RQ2: What sources can be used to build a CTI for AI knowledge base? How reliable are these data sources?}

We identified three main categories of sources for building CTI for AI knowledge bases: vulnerability-oriented, incident-oriented, and adversary-oriented sources. There are also specialized datasets, such as prompt injection datasets and malicious model repositories. Their quality varies: prompt injection datasets have variable quality, with recommended datasets (Qualifire, Prompt Injection Attack Dataset) being reliable for research, while many others have limitations in labeling or scope. Malicious model repositories have few documented cases (e.g., ReversingLabs investigations, NSFOCUS reports); they are reliable for some of the cases, but their coverage is very limited. Overall, current sources provide a foundation but remain incomplete. MITRE ATLAS and AIID are the most mature and reliable. AVID shows promise but needs more contributors. Many specialized datasets have quality issues.

\paragraph{RQ3: How would a CTI for AI knowledge base benefit tools designed to protect AI?}

A CTI for AI knowledge would provide important support for AI protection tools. By storing signatures and patterns of known malicious models, datasets, and attack techniques, tools could scan AI systems before deployment or investigate suspicious behavior by comparing it to past incidents. This helps identify likely causes and apply proven fixes, as enabled by resources like the AIID incident database and CSET taxonomy. Tools could also find which vulnerabilities affect specific model types, frameworks, or deployment scenarios, using repositories such as AVID. Encoding adversary TTPs from MITRE ATLAS allows monitoring for reconnaissance, data poisoning, or evasion attempts, similar to how endpoint detection and response (EDR) tools use ATT\&CK. With deep hashing and fuzzy matching techniques, even previously unseen malicious models can be detected if they are similar to known threats. Finally, structured taxonomies like AVID, CSET, and GMF help automatically sort incidents, assess severity, and guide response actions. In essence, a CTI for AI knowledge works for AI security tools much like traditional threat intelligence feeds do for network security, giving the context needed to detect, analyze, and respond to threats effectively.

\paragraph{RQ4: How to measure the similarity between collected IoCs and potentially malicious AI artifacts? How to effectively query the CTI for AI knowledge base?}

We identified several approaches for measuring similarity and querying AI-specific IoCs. Deep hashing and fuzzy hashing transform models, datasets, and features into compact codes, enabling fast comparison even for modified assets. Techniques from malware analysis, such as TLSH or LZJD, can also be adapted for AI assets. SCH preserves global semantic information, improving similarity detection. Practical CTI queries can combine exact lookups (hashes, repository names) with similarity searches using deep or fuzzy hashes. Features from model architectures, weights, or dataset profiles can be hashed for fast retrieval, and similarity scores from multiple methods can be aggregated to improve accuracy. Query strategies may include behavioral searches, contextual filters, and temporal analysis to track emerging threats.
The main challenge is balancing speed with detection accuracy. Efficient hashing enables real-time scanning, while richer similarity measures support deeper investigation, making a CTI system versatile and effective.

\section{Future Work}

Future research may focus on defining novel IoCs tailored for AI systems. Current CTI methods mostly focus on traditional software and networks, but AI frameworks, models, and datasets have unique behaviors and vulnerabilities. It may also be worth investigating what signs indicate that an AI model has been tampered with. This could include unusual model outputs, poisoned training data, or suspicious model updates. Another direction is to design and test these AI-specific IoCs in practice, and to create a framework that helps security teams monitor and respond to AI threats.

\end{multicols}

\clearpage
\printbibliography

\end{document}